# Computational Aided Design for Generating a Modular, Lightweight Car Concept

A.Farokhi Nejad[1,2 *], M. pourasghar[2,3], S.Peirovi[2], M.N.Tamin[2]

[1] *Department of Mechanical and Aerospace Engineering, Politecnico di Torino, Turin, Italy.*
[2] *Department of Mechanical Engineering, Universiti Teknologi Malaysia, Johor, Malaysia*
[3] *Automatic Control Department, Universitat Polit`ecnica de Catalunya, Barcelona, Spain.*

___________________________________________________________________________________________________________________

**Abstract**

Developing an appropriate design process for a conceptual model is a stepping stone toward designing car bodies. This paper presents a methodology to design a lightweight and modular space frame chassis for a sedan electric car. The dual phase high strength steel with improved mechanical properties is employed to reduce the weight of the car body. Utilizing the finite element analysis yields two models in order to predict the performance of each component. The first model is a beam structure with a rapid response in structural stiffness simulation. This model is used for performing the static tests including modal frequency, bending stiffens and torsional stiffness evaluation. Whereas the second model, i.e., a shell model, is proposed to illustrate every module's mechanical behavior as well as its crashworthiness efficiency. In order to perform the crashworthiness analysis, the explicit nonlinear dynamic solver provided by ABAQUS, a commercial finite element software, is used. The results of finite element beam and shell models are in line with the concept design specifications. Implementation of this procedure leads to generate a lightweight and modular concept for an electric car.

*Keywords*: Electric vehicle, Crashworthiness, Lightweight design, Modular concept, Space frame, Structural integrity.

___________________________________________________________________________________________________________________

## 1. Introduction

Nowadays, we are facing serious ecological issues among which global warming and air pollution are of greatest attention. More than 45% of the fuel consumption in passenger's cars is related to the body weight [1]. Reducing the weight with optimum design shows a great potential for solving this problem. Some studies conducted in this area show that taking the sophisticated approach of lightweight structural design can decrease fuel consumption significantly, leading to improving the aforementioned global issues [2,3].

Lightweight design is a vital aspect where mass is a critical design factor. In order to increase the driving comfort, safety and reducing the fuel consumption, the lightweight approach enables manufacturers to develop the products functionally [4,5]. To build a lightweight body car using high strength steel (HSS) [6,7], aluminum alloys [8] and composite materials have been proposed for example in [9,10]. However, the cost of the final component made by special non-steel types of materials is one of the obstacles that persuade manufacturers to employ high strength steel instead of the other materials [11]. Since some parts of body structure have low stress during the testing procedure, these parts can be replaced with lighter or cheaper materials. This approach called multi-mixed material that it can be used when the mass production is taken into account [4]. Also, the manufacturing process and formability of materials are the key points for obtaining the lightweight structures. In the mass production and especially for the automotive industry, the forming process inducing for example work hardening or material orientation offers possibilities to reach lighter components [12,13]. In addition, however, using optimal design has been considered as an another option to design a lighter car. Shape, compliance and mass optimization as well as genetic algorithm and neural network methods have been used to optimize the performance of car body and its component [14,15]. However, in some cases the methods such genetic algorithm and neural network for industrial application were not successful.

Crashworthiness assessment of the body car is a crucial issue that the manufacturers are concerned about and in recent years the regulations and consumer tests about the crashworthiness efficiency are becoming more challenging. The body structure plays the most important role to absorb the energy of the crash for the passenger cars. Therefore, in order to obtain lightweight vehicle regarding high crashworthiness efficiency, shape optimization was utilized in car pillars as proposed by [7,16]. The space frame chassis can be considered as one of the options to create a concept model that it can be optimized when modularity is taken into account.

Based on the definition of Original Equipped Manufacturer (OEM) standard for automotive industry, the modularity is "a group component, physically close to each other that both assembled and tested outside of facilities and can be assembled very simple on to a car". Furthermore, two different ap-proaches for modularity can be implemented in automotive industry namely modularity in design and modularity in assembly [17]. Regarding to design the modular concept, some manufacturers have introduced modular concepts to the market although, there is no standard for modularity approach when small number of production is needed [18]. One of the advantages of modularity is functionally-based optimization process. In the other word, the component of each module can be redesigned and optimized based on their application. For instance, shape optimization algorithm was used to evaluate the structural integrity of modular components from the same product family [19].

[*]Corresponding author. Fax.: +39-011-0906999
E-mail address: ali.farokhi@polito.it

In addition, using modal analysis and studying the natural frequency of each module as well as of the whole structure can be considered as a guideline for designers to better understand about the structural stiffness. Furthermore, study on the natural frequency of every module can not only bring the lightweight chassis but it would also yield to make much higher level of comfort and ride handling in the final design [20].

In order to create a new concept for academic purpose, it is difficult and expensive to access industrial tools instruments. Moreover, the car manufacturers are using different software and tools to generate a new concept. However, when a researcher or student team needs to design a specific prototype should consider different aspects such as structural integrity, dynamic response and crashworthiness. In this paper, a simple methodology is introduced in order to design a light weight and modular car body prototype. Following this methodology brings a fast response for studying the overall behavior of a car body structure regarding the small scale production.

## 2. Design Process Flow

In order to identify the design specifications, three main factors are considered for designing the prototype model: structural stiffness, modal frequency and crashworthiness. Table 1 gives the targets, which are defined here in designing a sedan electric car model. Based on the design targets, some general specifications such as wheelbase, width track dimensions, total weight and body frame weight are considered and subsequently, the computer aided design (CAD) model is created by using the commercial software SolidWorks. In order to reduce production costs, the conceptual space frame is designed with majorly rectangular profiles connected via modular joints as shown in Figure 1. Figure 2 indicates the four main modules that generate the space frame platform of a sedan car. The first module is the deck module with all the battery pack, electrical components, and the area for the passenger seats. The battery pack is placed between right and left seats longitudinally. At this place the air flow from the below the car can be helpful to increase the rate of heat transfer from batteries. Increasing the length of the two longitudinal beams in the deck module can change the wheelbase of the car; hence, the interior space of the care is increased. The second module is the front module that is responsible to protect passengers from frontal crash.

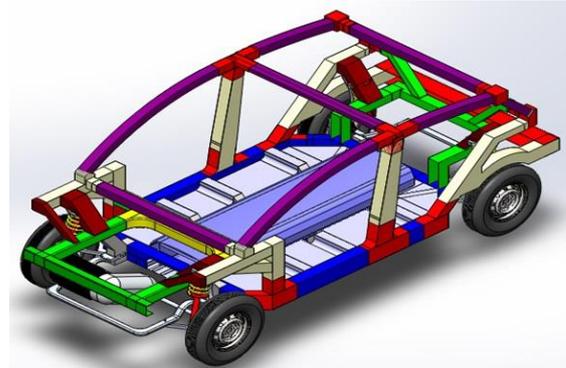

Figure 1. The conceptual design of sedan electric car space frame.

Table 1. Design specification for sedan electric car.

| Classification | | Target |
|---|---|---|
| Wheel base | | 2700 mm |
| Width track | | 1650 mm |
| Total weight | | < 1000 kg |
| BIW total weight | | < 250 kg |
| Natural frequency | | > 38Hz |
| Bending stiffness | | > 10 kN/mm |
| Torsional stiffness | | > 12 kN·m/deg |
| Frontal crash | Maximum intrusion | 110 mm |
| US-NCAP | Maximum deceleration | 30 g's |
| Rear crash | Maximum intrusion | 145 mm |
| FMVSS-301 | Maximum deceleration | 16 g's |
| Lateral crash | Maximum intrusion | 285 mm |
| FMVSS-214 | Maximum intrusion velocity | 9 m/s |
| Roof crash | Maximum intrusion | 127 mm |
| FMVSS- 216 | Max velocity | 5 mm/min |

However, holding motor, gearbox and the front suspension system would be the second function of this module. Furthermore, the third module is the rear module that is tasked to protect the passengers from the rear impact. Although in engine cars the first function of this module is to protect the fuel tank, in case of an electric car, this function is neglected [21]. Protecting the battery pack is a common task between first, second and third modules from different impact scenarios. In addition, holding the rear suspension system of the car is the other task of the rear module. Finally, the fourth module, the roof module, is the module with the role of increasing the body strength and stiffness. Moreover, the fourth module is used to connect all modules together. Increasing the strength against roll over and roof crash is the other specification of this module.in addition, the roof module is tasked to connect all three main pillars in order to provide desire torsional stiffness for the frame.

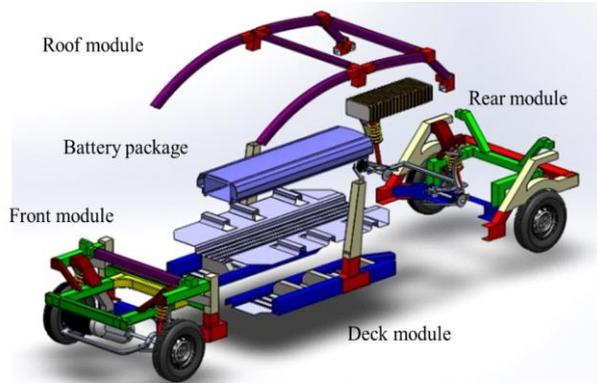

Figure 2. The four major modules of the space frame.

Modularity can be reached by using the modular connecting joints that attach the mentioned four modules together [18]. Additionally, modularity can increase the rate of production as well as production simplicity. For instance, in order to convert a sedan car to a hatchback, more than 80% of the components would be the same and for changing from sedan to sport utility vehicle (SUV), the space frame would consist of 60% of identical parts with sedan car [22]. Therefore, creating a reliable sedan car concept model can provide a good basis in modeling of the other class of this electric car. After defining the design specification and modularity consideration, the concept model should be remodeled by the finite element analysis (FEA).

Concerning computational time and for fast initial assessments of mechanical responses, it is necessary to develop a model with one-dimensional beam elements. Firstly, a conceptual design based on design targets is generated. Before starting a complex model, it is necessary to insure that the concept is strong enough against static loads. In order to achieve this goal, a beam model based on the initial concept dimensions is created. As the first test, a free body modal analysis to reach the frequency target (upper than 38 Hz) was performed. It is obvious that the first try is not the best design. To reach the optimum design every modules of the car is simulated separately and the highest frequency is picked as the best design for that module. The reason of this method is for reducing the computational time and avoiding random response. The tests are performed regarding the design constraints such as dimensional constraints, position on the joints, and the weight of each module. Considering this method helps to reach the higher natural frequency after 5 to 6 tries and each try takes less than thirty seconds. When the best response of each module is obtained the whole model is reassembled and natural frequency of the whole system is evaluated. If the target is reached the design can be considered as the final design otherwise the weakest module should be modified. At the end of this optimization loop the final conceptual design is obtained that is shown in Figures 1 and 2. The present model can be employed for evaluation of the bending stiffness and the torsional stiffness.

## 3. Beam Model Analysis

ABAQUS commercial code with B31 element type is used to model the beam space frame. In all tests, HSS material properties are assigned to the model, the cross-sections are rectangular between 40-70 mm and the thickness of material is varied between 0.7 mm to 1.2 mm. In addition, the backbone beams, pillars and longitudinal shot guns were defined to be thicker than the rest of components.

### 3.1 Modal analysis of the beam model

To assess modal frequency, no constraints are assigned and the frame is free [23]. For evaluation of the natural frequency of the system the Lankosz solver is used. The first significant mode shape is the most critical one, which should not meet the idle motor frequency. Figure 3 demonstrates the first and second mode shapes, which are the torsion and bending modes. The second mode is not the pure bending mode and it is the bending and torsion mixed mode. It can be said that at this frequency range the bumper has resonance effect that it is a transient mode. By increasing or decreasing the speed this phenomenon can be removed from the structure.

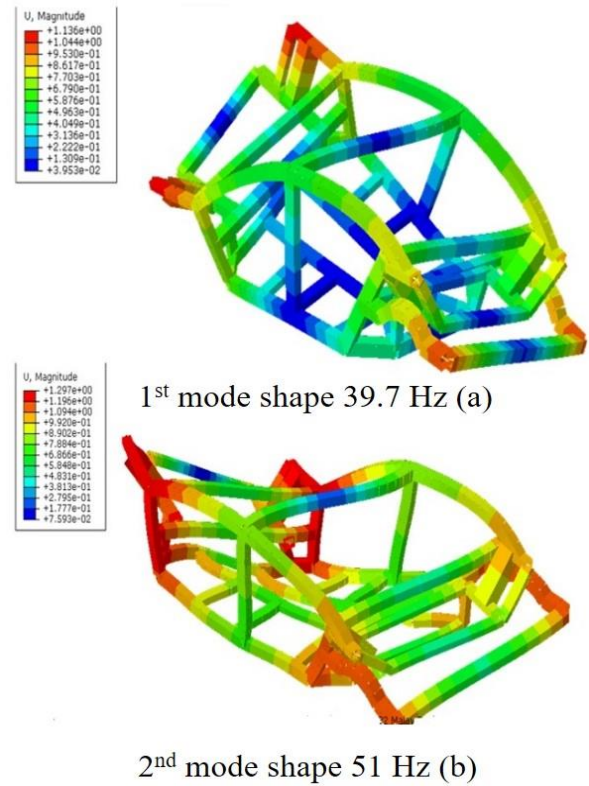

Figure 3. The significant modal shapes of the beam model. First mode(a) and second mode(b) of modal analysis

## 3.2 Bending stiffness evaluation of the beam model

To apply the bending stiffness test as the second test of this study, the boundary conditions were applied to the beam model. All four springhouses are constrained in three degree of freedom (Ux,Uy,Uz) and static loads are applied to represent the passenger's weight, battery pack and electric devices, which are distributed uniformly over several points [22]. In this case, 36 points are used to provide uniform distribution of the 5036 N load to the structure that is shown in Figure 4a. these number of points are related to the place of seats, batteries, motor, power train system, spare tire and the weight of final body. The elements size that is used in this test is equal to 15 mm. Figure 4b indicates the maximum vertical deflection. Dividing the total applied loads to the maximum vertical deflection determines the bending stiffness of the car structure.

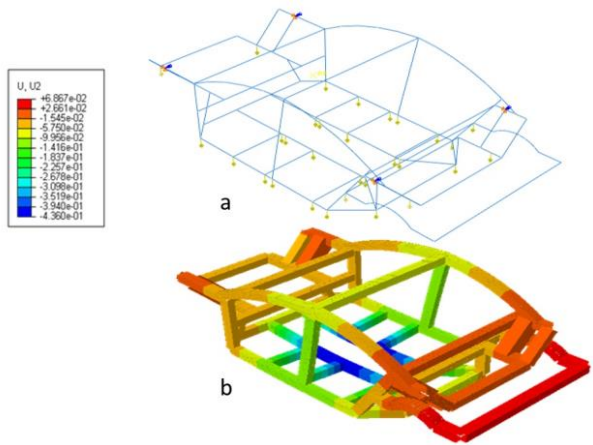

Figure 4. The maximum vertical deflection from bending stiffness test

## 3.3 Torsional stiffness evaluation of the beam model

Torsional stiffness is the third simulation that is performed on the space frame beam model. The torsional forces are imposed on front Springhouses as a torque and the rear Springhouses are constrained in three degree of freedom (Ux,Uy,URz). To determine the torsional stiffness, the following equations are suggested by Tebby et al. [24] where the torsional stiffness is represented by KT, F indicates the vertical force and B stands for the track width. Moreover, $v_d$, $v_p$, $\varphi_d$ and $\varphi_p$ are representing the vertical displacement and angular deflection of the front suspension positions around longitudinal axis, respectively.

$$K_T = \frac{T}{\phi_{ave}} = \frac{F \times B}{(\phi_d + \phi_p)/2}$$

$$\phi_d = \tan^{-1}(\frac{v_d}{B/2}) \quad (1)$$

$$\phi_p = \tan^{-1}(\frac{v_p}{B/2})$$

Since this space frame has a symmetric geometry, equation set (1) can be written as:

$$K_T = \frac{F \times B}{\tan^{-1}(\frac{U_{\max}}{2B})} \quad (2)$$

Where $U_{max}$ is the maximum vertical displacement at suspension position that can be identified in Figure 5.

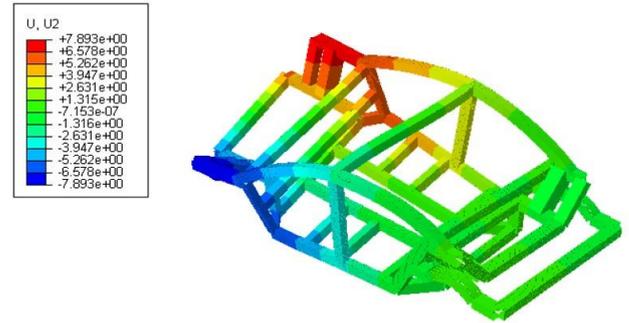

Figure 5. The maximum vertical deflection due to torsional stiffness test

## 4. Shell Model Analysis

Having completed the beam model and obtaining the specified targets, the model can be converted to a 2D element model with more complex element formulation in order to demonstrate both the mechanical behavior and the crashworthiness [25]. The two dimensional shell element with four nodes (S4R) is used as the element type [26]. Similarly, the shell model is generated by using the ABAQUS software based on the CAD concept and this model covered all the simulation tests such as static tests, modal frequency and crashworthiness analysis tests. For this reason, a mesh convergence study is performed to get more accurate results for different kind of analysis. Moreover, the energy balance study is considered for the crash tests analysis.

### 4.1 Material properties

The material properties of dual phase HSS (DP600) from our previous research [26] is taken into account. To consider high strain rates, the empirical Johnson Cook model is used:

$$\overline{\sigma} = A + B(\overline{\varepsilon_{pl}})^n ][1 + C \ln(\frac{\dot{\varepsilon}_{pl}}{\dot{\varepsilon}_0})][1 - (T^*)^m] \quad (3)$$

Where A, B, C, m, n are material constants that are extracted below transient temperature (T*). In this study the temperature gradient is neglected. Table 2 indicates the DP600 Johnson Cook characteristics. For the static and modal tests, the elastic properties of the DP600 are used.

Table 2. The DP600 Johnson Cook parameters [26].

| parameters | A (MPa) | B (MPa) | C | m | n |
|---|---|---|---|---|---|
| value | 350 | 902 | 0.014 | 1.23 | 0.189 |

### 4.2 Modal analysis of the shell model

Figure 6 illustrates the first two significant modal frequencies, that are related to torsion and bending modes, respectively. As a result of adding joints to the shell model regarding perfectly bonding, this model is observed to be stiffer in comparison with the beam model. However, adding these joint increases the weight of structure that it is cause of reduction of natural frequencies in this model rather than the beam model. The results of this model shows that the target is achieved. The first mode shape of both models show that the space frame is weak against torsional force.

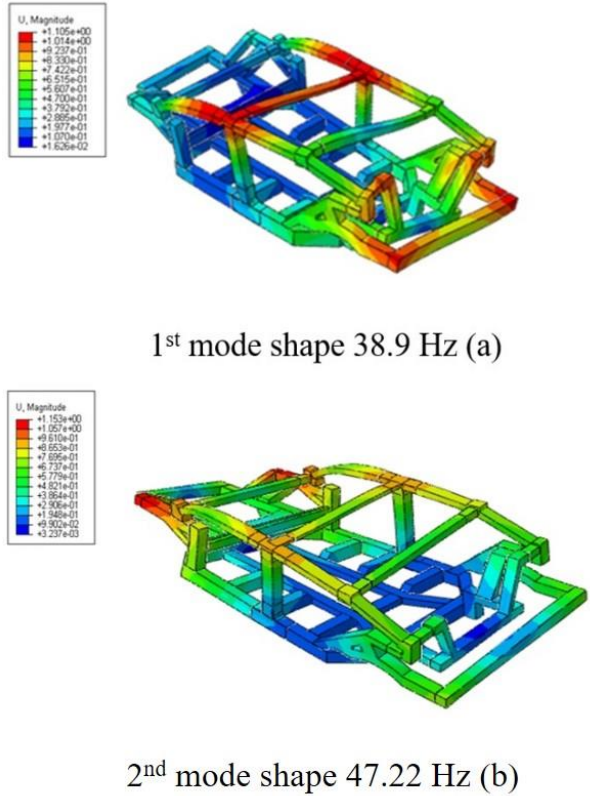

Figure 6. The modal frequencies from the shell First mode(a) and second mode(b) of modal analysis

### 4.3 Bending stiffness evaluation for shell model

Figure 7 indicates the vertical deflection of the bending stiffness test for the shell model, in which the largest element size is assigned to be 15 mm. In this simulation, all springhouses are constrained in three degrees of freedom same as beam model and all loads are distributed at the same previous points from the FE beam model.

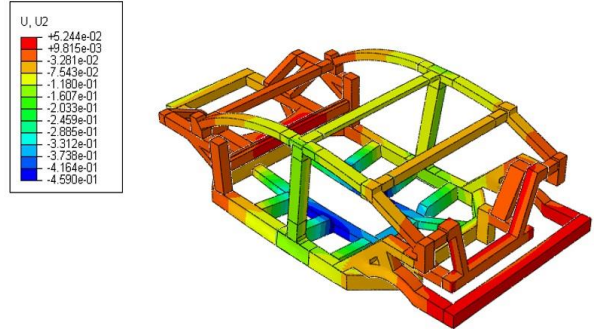

Figure 7. The vertical deflection in bending stiffness test from the shell model

### 4.4 Torsional stiffness evaluation for shell model

Figure 8 shows the vertical displacement obtained from the torsion test. In this simulation, rear suspensions are fixed in six degrees of freedom and a torque is applied on the front springs. Substitution of the value of the vertical displacement in (2) yields the torsional stiffness of the model. It can be expected that adding joint and other sheet plates to the final structure as exterior closures will increase the torsional stiffness of the body car.

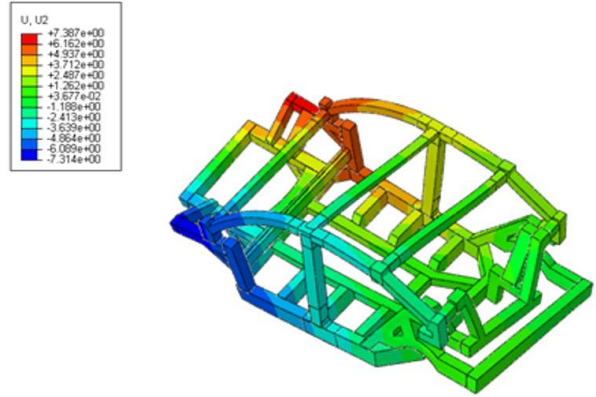

Figure 8. The vertical displacement resulting from the torsion test

## 5. Crashworthiness Analysis

In order to follow the related standards for crashworthiness analysis the new car assessment process (NCAP) standard test for frontal integrity and Federal Motor Vehicle Safety Standard (FMVSS) are considered for the lateral, back and roof crashworthiness assessments. In addition, for obtaining more accurate result, the total weight of the car is assigned to the frame's center of gravity. ABAQUS/Explicit is employed to simulate the crash tests.

### 5.1 The frontal crash test

Currently, the full width frontal crash test has been paid attention by car manufacturers due to the test reliability. In the better word when a car passes this test successfully, it means that the structural integrity is appropriate for all different front side impact scenarios. Figure 9 shows the deformed shape of a space frame for frontal crash simulation, that is performed as defined for the US-NCAP requirements. In this simulation, the car collided with a rigid barrier directly by 55 km/h speed within 90 milliseconds. Elements with size of 10-mm are allocated to the bumper, longitudinal beams and upper rails, whereas the elements of the pillars and front passenger cabin have a of 30-mm size; however, the remaining parts are assigned with 40-mm size of elements.

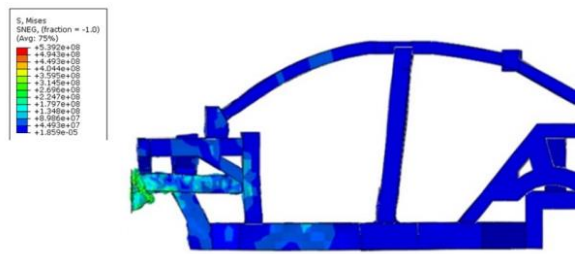

Figure 9. The frontal crash simulation US-NCAP within 90 ms.

### 5.2 The lateral crash test

Lateral crashes consist more than a quarter of number of deaths for passenger vehicle car around the world [27]. Passengers protection subjected to the side impact is a challenging issue due to the little space for energy absorption. Recently, using side airbags are taken into account by car manufacturers; however, the structure strength and energy absorption plays the key role to protecting the occupants. Figure 10 shows the deformed shape of the structure under lateral crash simulation condition. Based on FMVSS- 214 lateral crash standards, a deformable barrier is colliding to the frame at the speed of 50 km/h and an angle of impact of 27° within 90 milliseconds. In this test, the size of the elements of the side part is assigned 15 mm and the remaining parts are considered 40 mm.

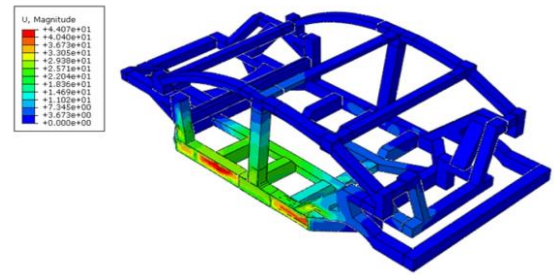

Figure 10. The lateral crash simulation based on FMVSS-214 test standards

### 5.3 The rear crash test

The main purpose of performing the rear crash test is protection of fuel tanks for combustion engine cars to avoid the post-crash fire. However, in electric car the main task for this test is protection of passengers from rear impact. In this case due to using lightweight design approach the energy absorption from rear side should be considered. Figure 11 illustrates the maximum deformation due to rear impact. In the rear crash test, a rigid barrier collided with a velocity of 50 km/h to the rear bumper directly within 90 milliseconds. The size of elements in the rear bumper and longitudinal beams were 15 mm and the other parts were 40 mm.

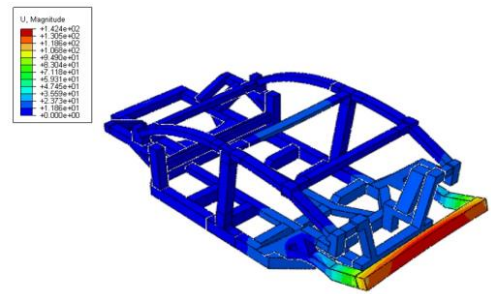

Figure 11. The rear crash test, accomplished based on FMVSS-301 standards in 90 ms.

### 5.4 The roof crush test

The number of casualties from rollover crashes show that these kind of events are serious destructive for the passengers. The evidences show that the major damage usually includes Pillars and roof deformation [28]. Whereas, the roof test crashworthiness assessment is crucial for designing the new car. Figure 12 shows the maximum deflection from the roof crash test at the end of simulation. To simulate this test, a 14700 N load, i.e. 1.5 times larger than the car's total weight, is applied on the roof of the vehicle by a rigid plate. The applied load velocity was 5mm/min that it can be considered as the quasi static loading. The angle of contact between the plate and the roof were considered 5° and 25° along X and Z directions, respectively. In addition, the lower rocker is constrained in six degrees of freedom.

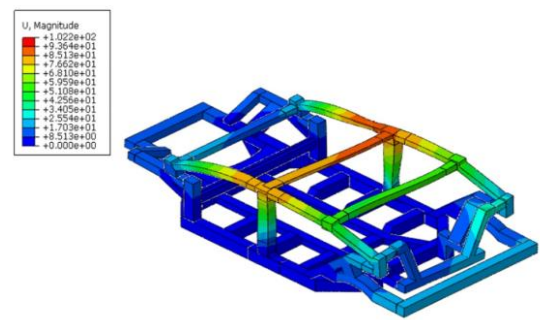

Figure 12. The roof crash test FMVSS-216 within 90 ms.

## 6. Summary of Results

The following section gives a summary of the results obtained by the two FE models, namely the beam and the shell model simulation results. As the total weight of the space frame is 167 kg, it can be considered a lightweight body car in this class of automotive. Table 3 compares the results of a beam element and a shell element for static test including bending stiffness, torsional stiffness, and modal frequency simulations.

Table 3. The results of beam and shell models for structural simulation

| Static tests | Target | Beam model | Dev (%) | Shell model | Dev (%) |
|---|---|---|---|---|---|
| Bending stiffness (kN/mm) | 10 | 11.53 | 15.30 | 10.96 | 4.47 |
| Torsional stiffness (kN·m/deg) | 12 | 11.67 | -2.75 | 12.20 | 1.66 |
| Modal analysis (Hz) | 38 | 39.70 | 9.60 | 38.34 | 0.89 |

The table above shows that all tests met their expected targets except the torsional stiffness of the beam model, which is due to the nature of the space frames. Since space frames are weak innately, adding joints is required to improve the stiffness. Furthermore, it can be expected that adding sheet floor and other body closures increases body stiffness [29]. Therefore, it is required to repeat the previous tests after installation of all the body components. The deviation between two models and the design targets show that by increasing the order of elements the result deviation will be closed to the target. However, all simulations except for beam torsional stiffness, the deviations are positive and it increases the structure integrity. The maximum error between two models is around 11% and it is due to adding joints to the shell model. In this study the joints are modeled as perfectly bounded however, in real tests it can be expected that the test values should be lower than in the numerical models. Therefore, it can be interpreted that the shell model has more precise results. Table 4 (frontal crash) shows the results of frontal crash simulation based on US-NCAP standards, as well as the preferred targets of these simulations. The maximum deceleration and intrusion were measured from the driver foot place. In the real test these data are collected from different position e.g. the head and the feet of dummy driver, A and B pillars. According to this table, the obtained values are all well below the defined maximum standards, making them acceptable in terms of matching US-NCAP standards.

Table 4. The crashworthiness assessment of for different crash tests.

| Type of crash test | Physical identification | Target of test | FEA result | Min dev (%) |
|---|---|---|---|---|
| Frontal crash US-NCAP | Maximum intrusion (mm) | 110 | 82 | 16 |
| | Maximum deceleration (g's) | 30 | 25 | |
| Rear crash FMVSS-301 | Maximum intrusion (mm) | 145 | 142 | 2 |
| | Maximum deceleration (g's) | 16 | 8.5 | |
| Lateral crash FMVSS-214 | Maximum intrusion (mm) | 285 | 44 | 28 |
| | Maximum intrusion velocity (m/s) | 9 | 6.45 | |
| Roof crash FMVSS-216 | Maximum Intrusion (mm) | 127 | 102.2 | 19 |
| | Max velocity (mm/min) | 5 | 1.5 | |

The lateral crash simulation results, which are within the appropriate range and are all below the maximum allowed values, are presented in Table 4 (Lateral crash). This simulation is conducted in conformance with US-FMVSS 214 side crash standards. The maximum intrusion and hence, its velocity were measured from the longitudinal beam between A and B pillars and near to the driver position. This test shows the integrity of space frame from side crash. Unlike the frontal test the barrier made by deformable elements thus, some parts of impact energy are dissipated on the barrier. It should be mentioned that, by installation of the doors, the plastic deformation will be increased and the deceleration time decreased. In other words, the minimum deviation percentage by adding the other components can be decreased. In this study, the target number of all the crash tests are taken into account from the testing standard for the car body. Therefore, it can be expected that the results of the final body car will be different with the body space frame. However, evaluation of the space frame crashworthiness brings a desire estimation for the final design integrity. As explained before, the rear crash simulation is carried out based on US FMVSS-301 test standards. Table 4 (Rear crash) presents the maximum deceleration and intrusion from the crash test, where both targets were met. The maximum intrusion is occurred near to the intersection area between C pillar and rear longitudinal beam. To examine the integrity of space frame's roof, according to US FMVSS 216, a roof crash simulation is performed. From Table 4 (roof crash,) it can be seen that the target of this simulation is achieved. Figure 13 illustrates the deceleration of the vehicle structure during simulation time. The comparison between three high impact tests: frontal, rear and lateral show that the maximum energy transfer to the car with the full width frontal test and by absorption of the energy with plastic deformation the deceleration becomes zero. In the lateral test due to the angle of attack the crumpling zone consists of a bigger area and the peak point of deceleration is not at the first peak load. In addition, the first and second deceleration peak point in rear test are related to the initial impact and bending of the rear bumper respectively. The simulation time for three tests were considered 90 ms although, the rear and lateral tests were finished after 75 ms. The mentioned standards for the crash tests in this study are given for full body test and in the real test the elastic rebound can be seen; however, in the frame crash test it can be expected that the results and overall behavior should have slightly different with final design.

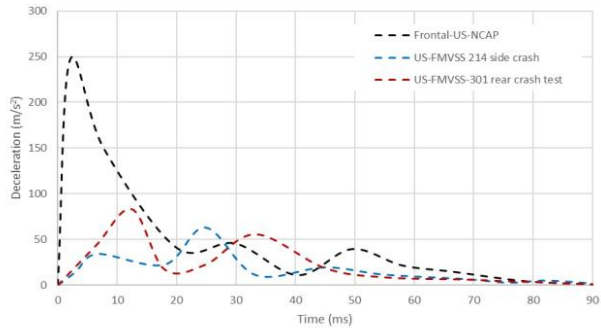

Figure 13. the vehicle deceleration subjected to frontal, rear and lateral crash tests within 90 ms.

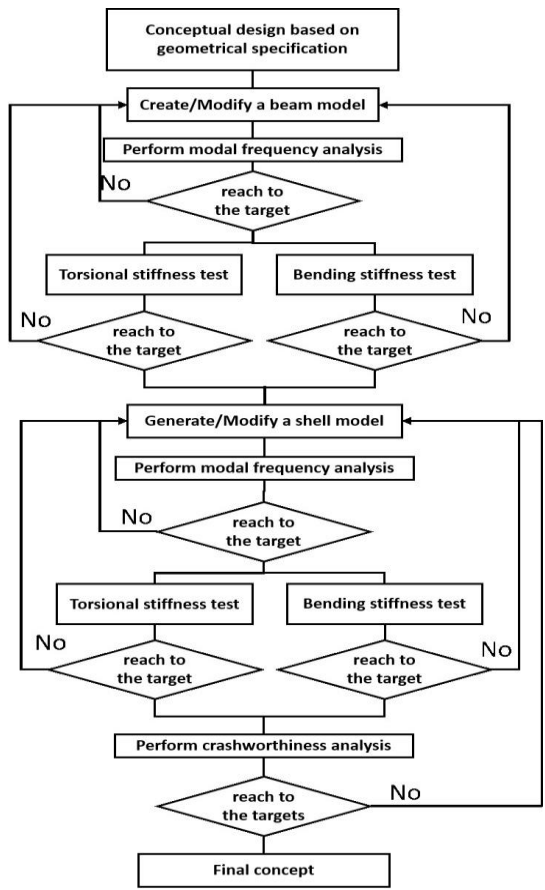

Figure 14. the flowchart of the design of a conceptual space frame car.

Figure 14 presents the design flow of a conceptual space-frame regarding static and crashworthiness tests. In order to design a lightweight and modular concept, the first design based on geometrical specification is proposed. A beam model with respect the concept is generated and the static tests are performed on it. After passing the design targets a more complex model with shell elements and consideration of modular joints is generated and the previous tests are repeated on the shell model. Having passed the targets, the model is used for the crashworthiness assessment. Therefore, the final concept can be introduced after finishing the crashworthiness evaluation. The shell model is able to modify or change every module's component in terms of modularity and lightweight approach.

## 7. Conclusion

In this paper, a methodology to design a lightweight and modular space frame chassis was developed. The DP-600 high strength steel with improved mechanical properties was employed as the body material with the purpose of reducing the weight. To predict the performance of different components, the finite element analysis was utilized with both beam and shell models as the first model is capable of providing rapid responses in structural stiffness simulations and the shell model can predict more complex behaviors such studies about the structure's crashworthiness. Implementation of this procedure leads to generate a lightweight and modular concept for a sedan electric car.

The results show the feasibility of a conceptual design as, the results of beam and shell model were higher than expected specification. In addition, application of the HSS increased the integrity of the space frame dramatically by decreasing body weight. Therefore, optimization is needed for reducing the space frame weight. The proposed design flow can be used for accelerating the design procedure and reducing the cost of design. However, further research is recommended to study the modular joints, which requires the use of multi-mix material.

## Acknowledgment

All the simulations in this work was performed in high performance computing (HPC) center in University Technology of Malaysia.